\documentstyle[preprint,prd,aps,floats,tighten,epsf]{revtex}
\begin{document}

\preprint{\vbox{\hbox{UCD 2000-3}}}

\title{Chiral condensate in the quenched Schwinger model}
\author{ Joe Kiskis}
\address{ Dept. of Physics, University of California,
Davis, CA 95616}
\author{ Rajamani Narayanan}
\address{ American Physical Society, One Research Road, Ridge, NY 11961 }
\maketitle 

\begin{abstract}
A numerical investigation of the quenched Schwinger model on the lattice
using the overlap Dirac operator points to a divergent chiral condensate.
\end{abstract}

\section{Introduction}

Use of quenched QCD as an approximation to the full theory depends upon a good
understanding of the regions of parameter space where the quenched theory
differs in important ways from the full theory. For the case of the chiral
condensate $\langle \bar\psi\psi \rangle$, there may be qualitatively different
behaviors for sufficiently small quark mass. Whereas the condensate is expected
to be finite in the full theory, there are theoretical arguments~\cite{Sharpe}
and some
initial numerical indications~\cite{QCD} that it diverges in quenched QCD. 
Also numerical analysis~\cite{Sharan} of an instanton
gas model shows a divergence.
A careful
study, using a lattice Dirac operator that obeys chiral symmetry, to determine
the mass range in which quenched QCD is a good approximation to the full theory
has not been done.

Similarly for the Schwinger model, the full theory has a finite 
condensate\cite{ls}
but there are predictions\cite{Smilga,Durr} that it diverges in the quenched 
theory. Thus the
Schwinger model can be used to investigate the two-dimensional versions of
these questions concerning anomalies, topology, chiral symmetry, and
condensates and their impact on the relationship between full and quenched
theories. Although the Schwinger model is in most ways
much simpler than QCD, it does present some peculiar
difficulties of its own. 
Strong infrared effects, which are dynamical and nonperturbative in QCD, are
already kinematical in the lower dimension of the Schwinger model.
There is the possibility that infrared enhancement in the quenched case gives a
fermion spectral density that is divergent as the eigenvalue $\lambda$ goes to
zero and that there is a corresponding infinite condensate $<\bar\psi\psi>$.
We have investigated this issue numerically using the overlap Dirac operator to
describe the massless limit for the fermions and have found strong evidence for
these divergences in the quenched Schwinger model.

When stated in terms of the low eigenvalue behavior of the fermion
spectral density in the quenched Schwinger model, theoretical discussions 
have given a lower bound that is
finite\cite{Casher} and stronger one that is divergent\cite{Smilga}.
Some estimates\cite{Smilga,Durr} that are not bounds have suggested 
a form diverging exponentially in the volume $e^{cg^2V}$. 
The discussions in Refs.~\cite{Smilga,Casher} are given in
terms of the eigenvalue shifts of the would-be-zero modes associated
with subregions of the whole two-dimensional volume $V$. With larger
shifts \cite{Casher} due to interactions with other subareas, the
spectrum is flat at small $\lambda$ in the infinite volume
limit. Smaller shifts \cite{Smilga} leave the would-be-zero modes
concentrated near the origin so that the spectral density there
diverges as $V \rightarrow \infty$. Other arguments \cite{Smilga,Durr} for
the form of the divergence proceed along different lines, but the
implication for the spectrum is that the lowest eigenvalues are
exponentially small in the volume with a corresponding exponentially
large spectral density and condensate.

The data that we present here covers a range of lattice sizes from
$8^2$ to $32^2$. The full spectrum of the overlap Dirac operator was
calculated in gauge backgrounds from the Wilson action at several bare
couplings. 
The next section discusses the lattice formalism used in this paper.
The third section discusses the physics issues in more detail. 
The fourth section
gives our numerical results. In addition to the spectrum itself, there are
measures of its behavior including $<\bar\psi\psi>$ and the distribution of the
lowest eigenvalue.
The last section contains a summary of our
results and some concluding discussion.

\section {Lattice formalism}

Since we are interested in studying the small mass region and the
massless limit, we need to work with a lattice Dirac operator that
respects chiral symmetry. We will use the overlap Dirac operator
for our numerical study. It has the form\cite{Herbert}
\begin{equation}
D={1\over 2} \Bigl[ 1 + m + (1-m)\gamma_5\epsilon(H_w) \Bigr ]
\label{eq:over}
\end{equation}
with $H_w$ being the hermitian Wilson Dirac operator in the
supercritical region and $0 \le m \le 1$ is the bare fermion mass.
The hermitian overlap Dirac operator $H=\gamma_5 D$ has paired
non-zero eigenvalues. The topological zero modes are chiral
and have partners with unit eigenvalue and opposite chirality.
In a fixed gauge field background\cite{EHN},
\begin{equation}
<\bar\psi\psi>= {|Q|\over mV} + {1\over V} \sum_{\lambda>0} 
{2 m (1-\lambda^2) \over \lambda^2(1-m^2) + m^2} .
\label{eq:pbp}
\end{equation}
The sum is over all positive non-zero eigenvalues of $H$,
$Q$ is the global topological charge,
and $V$ is the lattice volume. 

In the numerical calculation, we generate gauge fields
distributed according to the Wilson gauge action
\begin{equation}
S_g ={1\over g^2} \sum_p {\rm Re} U_p   
\end{equation}
with $U_p$ the product of U(1) link elements around a fundamental
plaquette and $g$ the lattice coupling constant. 
The fermions have periodic boundary conditions
\footnote{
This choice of boundary conditions is not as restrictive
as it seems since we only have one fermion. 
A gauge field
configuration can be multiplied by an arbitrary constant U(1) field
on each link in either of the directions without changing the gauge action.
Since all of these possibilities are included in the sum over gauge field
configurations, there is no real distinction between periodic 
and anti-periodic boundary conditions.
More generally all boundary conditions that are periodic up to a phase are
equivalent.
}.
For each choice of $g$ and $L$, we diagonalize $H_w$ in a fixed gauge field 
background
and form $H$ by first forming $\epsilon(H_w)$. We then diagonalize
$H^2$ in the chiral sector that contains topological zero modes, if any.
Since all computations are done in double precision, we know
the non-zero eigenvalues of $H$ to an absolute precision of $10^{-8}$.
In addition we know the exact number of 
zero eigenvalues of $H$ by counting the difference between
the number of positive
and negative eigenvalues of $H_w$~\cite{NN}.

\section{Physics issues}

In the multi-flavor Schwinger model, the classical U(1) chiral symmetry is 
explicitly broken by the anomaly, while the SU(N) chiral symmetry cannot be
broken in two dimensions.
The 't Hooft vertex $<\prod_i\bar\psi_i\psi_i>$
is not 
associated with an intact symmetry 
or Goldstone bosons, so it can and does have a nonzero value \cite{Hooft}.
In the quenched case, the exact zero modes of the massless Dirac operator
cause a divergence in $<\bar\psi\psi>$ in the
massless limit at finite volume. 
But as seen in (\ref{eq:pbp}), the divergence is of
the form $<|Q|>/(mV)$.
Since
$<|Q|>\propto \sqrt{V}$, it follows that 
\begin{equation}
\lim_{m\rightarrow 0}\lim_{V\rightarrow\infty} <|Q|>/(mV) = 0 .
\label{eq:global}
\end{equation}
This trivial divergence does not contribute in the case where
one first takes the thermodynamic limit and then takes the massless limit.
But this finite volume divergence does not appear
in the unquenched
Schwinger model. The zero modes of the Dirac operator in these
backgrounds cause a suppression of such gauge field configurations
when the fermion determinant is included as part of the
gauge field measure. 

The small eigenvalue behavior of the spectrum determines the contribution that
the second term in (\ref{eq:pbp}) makes to $<\bar\psi\psi>$. Thus the issues
to be numerically investigated are centered upon the small eigenvalue behavior
of the massless Dirac operator. The main question is whether the infinite volume
spectrum is flat as the eigenvalue $\lambda$ goes to zero or has a divergence
at small $\lambda$.

Consider the gauge field seen by the fermion.
The plaquette magnetic field of the quenched Schwinger model is ultra-local
with the field fluctuations on different plaquettes uncorrelated in infinite
volume. The plaquette angles are approximately gaussian distributed at weak
coupling. Thus the study of fermionic observables
in the quenched theory is best thought of as an investigation of a disordered
system~\cite{Casher}.

Let us begin the discussion with two much simpler examples.
For the case of free fermions on an $L \times L$ lattice with periodic
boundary conditions, the low-lying levels are 
\begin{equation}
 \lambda \approx [({2 \pi n_1 \over L})^2 + ({2 \pi n_2 \over L})^2 ]^{1/2}
% \lambda \approx [ (\case{2 \pi n_1}{L})^2 + (\case{2 \pi n_2}{L})^2 ]^{1/2}
\end{equation}
so that the level spacing is of order $1/L$, and the density of states per unit
volume is of order $\lambda$ at the low end.

Another simple case is a uniform magnetic field $B$, which
gives Landau levels. The level spacing is of order $B$, and the degeneracy of
each level is of order $BV$. With the scale of energy intervals larger than
$B$, the density of states is flat. As we will see later, the typical $BV$ is
$gL$ so that the average of $B$ over $V$ does get smaller with increasing 
volume. Thus the
Landau levels give a flat spectral density if the energy resolution is coarser
than $g/L$.

For the case at hand of particles with gyromagnetic ratio 2, there is a
cancellation between the paramagnetic magnetic moment interaction with the
field and the diamagnetic kinetic energy contribution that puts the lowest 
Landau level at exactly zero energy. These are the $BV/(2\pi)$ zero modes.

When the net flux is zero and all boundary conditions are periodic so that 
the vector potential can be put in the form
\begin{equation}
 A_\mu = \epsilon_{\mu\nu} \partial_\nu \phi ,
\end{equation}
there is also a pair of zero modes with opposite chirality. 
These have the form
\begin{equation}
 \psi_+ = e^{\phi} \left( \begin{array}{c}
                           1 \\ 0
                        \end{array} \right)
  \; \; \text{ and } \; \; 
 \psi_- = e^{-\phi} \left( \begin{array}{c}
                           0 \\ 1
                        \end{array} \right)
\label{eq:zero_twist}
\end{equation}                       

For the quenched Schwinger model the field is neither zero nor uniform.
As noted above, the
magnetic field is random with no plaquette-plaquette correlation between the
magnetic field on different plaquettes. Thus the variance increases as the 
area.
With the coupling $g$ small and $gR$
large, the flux through the area $R\times R$ is gaussian distributed with a
typical size of $gR$, so that the area average of the field strength is 
$B=g/R$.

\begin{figure}
\epsfxsize = 0.9\textwidth
\caption{Low end of the positive half of the
spectrum on a $L=32$ lattice on the special configuration
as a function of $R$ which defines the area of constant magnetic
field of flux $R$. 
Eigenvalues below $10^{-7}$ are not shown in this plot. Only for $R=3$ are all
the eigenvalues above this bound.
The lines connect eigenvalues at the same position in the ordering for each
$R$.}
\centerline{\epsffile{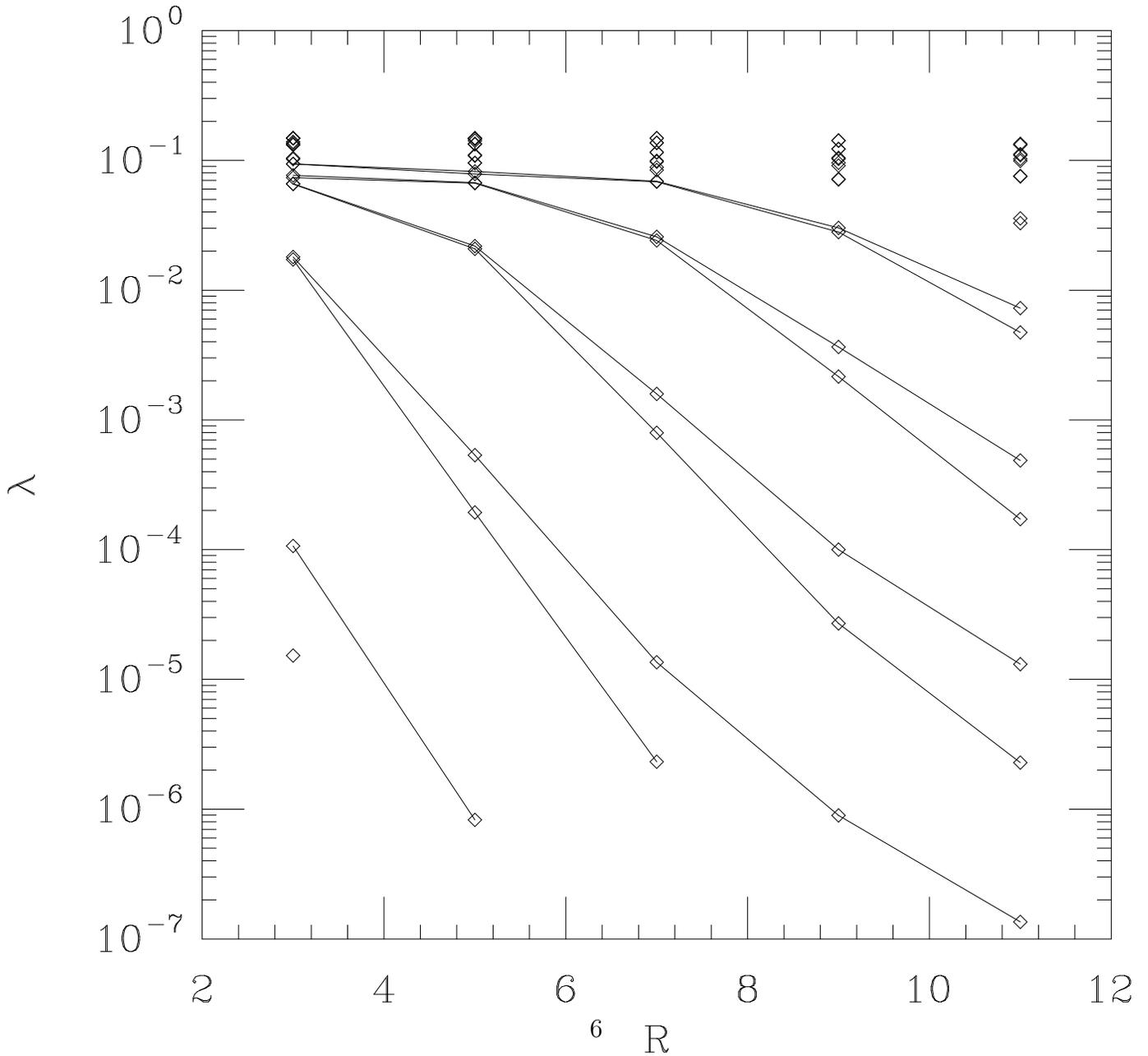}}
\label{fig:special}
\end{figure}

What is the fermion spectrum in that case? The index theorem tells us
that for a net flux $2 \pi f$ through the area $L^2$, there are $f$ modes
with zero eigenvalue.  For $f=0$ and all boundary conditions 
periodic, there are
two zero modes of the form above. But what else happens at the low end
of the spectrum? There are two suggestions for an answer in the
literature. The discussion of Casher and Neuberger~\cite{Casher}
begins by dividing the $L\times L$ volume into $R\times R$-sized
pieces with $gR$ and $L/R$ large. Considered in isolation, each of
these areas has of order $gR$ zero modes. It is then argued that the
effect of interaction between different regions is to shift the
eigenvalues of the zero modes away from zero, in such a way as to
produce a spectrum that is bounded below by one that is flat at small
$\lambda$.

With a similar approach, Smilga~\cite{Smilga} argues for a stronger
lower bound that gives a spectral density that diverges as $\lambda
\rightarrow 0$. His stronger result follows from using the fact that
the value of the zero mode wave function on the boundary that
separates a region with magnetic field from one with zero field is 
exponentially small
in the flux through the region.  Arguments using other methods in his
paper and in the paper by D\"urr and Sharpe~\cite{Durr} conclude that
the divergence is quite strong with a factor $e^{cg^2V}$. This would
be a consequence of modes with eigenvalues as small as the inverse of
that factor.

For a numerical test of the argument used by Smilga to produce the stronger
bound,
we construct a background gauge field configuration that has
two regions of size $R^2$ each with constant magnetic field 
and opposite net fluxes of magnitude $R$.
We 
study the low lying eigenvalues of the overlap Dirac operator 
and show that they go down exponentially with $R$.

On the $L^2$ lattice, fix two regions of size $R^2$ separated
by $(L/2-1,L/2-1)$.  The slightly off-symmetric separation is chosen
to avoid any accidental lattice symmetries. On one region of size
$R^2$, we make up a constant magnetic field of flux $R$, and on the
other region, we make up a constant magnetic field of flux
$-R$.  Elsewhere the field is zero. 
An initial numerical check with the field set to zero in one
of the regions, verified the presence of $R$ topological zero 
modes for
the overlap Dirac operator. Then returning to the case of interest
with the field in both regions, we calculated the spectrum again.  In
Fig.\ref{fig:special}, we plot the low end of the
positive half of the spectrum as a
function of $R$. 
The lowest few of these small
eigenvalues go down exponentially in $R$. 
We verified that this behavior
remains unchanged when small random perturbations are added  to
the link elements. The numerical results are
less restrictive than the theoretical arguments in that they do not
rely on a variational argument, which without further work, applies
only to a single mode on the lattice. Also all the lattice modes in
addition to the would-be-zeros are included. This confirms the crucial
point in the argument for the stronger bound of Ref.\cite{Smilga}. 
However, it does not
provide evidence that the spacings might be as small as $e^{-cg^2V}$.

Let us now consider the continuum limit. Given the remark above on the lack of
correlation in the field strengths, it is not possible to use a scale from that
in defining the continuum. However, there is another simple approach. For $g$
small, ask how large does $R$ have to be so that the typical flux through the
$R\times R$ area is of order one? Since the flux variance for a single 
plaquette is 
$g^2$ and the uncorrelated fluxes add, the variance for the area is $g^2R^2$.
Thus the typical flux through the area is $gR$, and $g$ sets an inverse length 
or energy scale. This happens to be of the same order as the scale for 
the unquenched
Schwinger model, in which the mass in lattice units is $g/\sqrt{\pi}$. 
We may hope to get a sensible continuum limit by measuring
continuum dimensionful quantities in units of appropriate powers of $g/a$. 
To get the
finite volume continuum limit, we will want to take $g$ to zero and $L$ to
infinity with $gL$ fixed. Lattice eigenvalues, $<\bar\psi\psi>$, and other 
quantities with continuum units of energy should be considered in ratios like 
$\lambda/g$ as $g \rightarrow 0$.

Finally, let us discuss the range of $g$ and $L$ where these
interesting effects might be seen. From two points of view, we can see
that $gL$ must be large. First if $L$ is fixed and $g$ is small, then
there are only perturbative effects from the gauge field, and the
strong infrared fluctuations cannot appear. Also if $gL$ is small,
then there are essentially no would-be-zero modes that could realize
the the physical pictures of \cite{Casher} and \cite{Smilga}. If $g$
is large, then the gauge field is very rough on the scale of a single
lattice spacing, and the continuum-based arguments above do not apply.
The smallest region that typically contains a unit of flux should be
several lattice units so that it is large enough for the fermion to
realize zero modes from the non-zero flux.  
Thus we must have $g$ small and $gL$ large, which means, of
course, that $L$ must be large.

\section{Numerical results}

\begin{figure}
\epsfxsize = 0.9\textwidth
\caption{Plot of $<\bar\psi\psi>$ with respect to $m$ at $g=0.4\sqrt{\pi}$
for four different lattice sizes}
\centerline{\epsffile{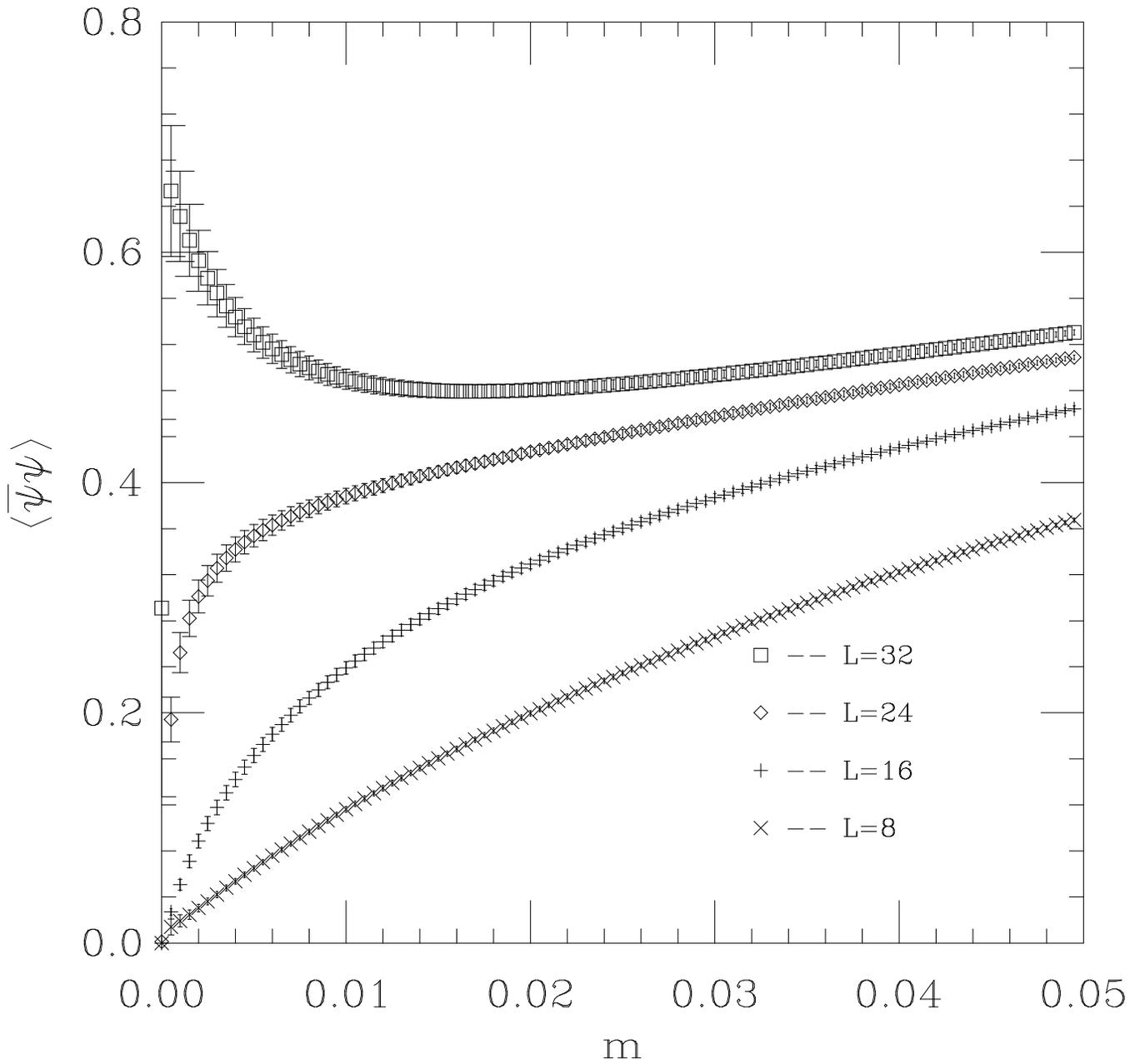}}
\label{fig:pbp1}
\end{figure}
\begin{figure}
\epsfxsize = 0.9\textwidth
\caption{Plot of the distribution of the small non-zero eigenvalues
at $g=0.4\sqrt{\pi}$
for four different lattice sizes}
\centerline{\epsffile{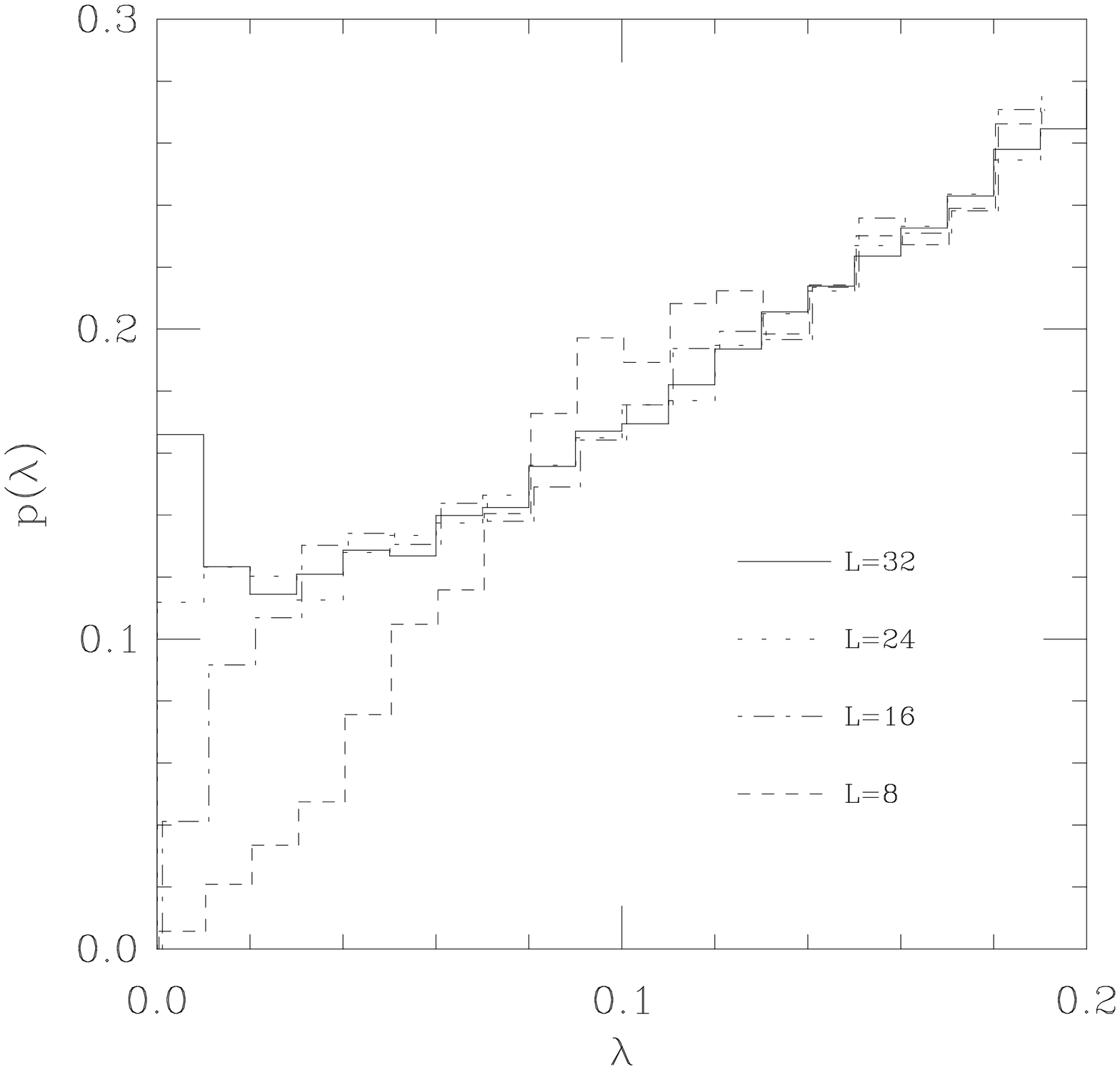}}
\label{fig:hist1}
\end{figure}

The numerical results described in this section give
substantial evidence that the infinite volume limit of the spectral density 
$\rho(\lambda)$ is indeed infinite for $\lambda \rightarrow 0$.
We will show this by computing the spectrum of the overlap Dirac operator
in U(1) gauge field backgrounds. 
The massless limit
is approached in the conventional way by adding a standard mass term.
We will also compute $<\bar\psi\psi>$
as a function of mass in finite lattice volumes and show that it grows
as the mass is lowered before it finally dives to zero 
as it must for $m=0$ and finite volume.
We will also show that the average value of the lowest eigenvalue
does not scale with the volume, nor does it fit predictions from
chiral random matrix theory.

Strong evidence for a divergence in the non-topological
piece of $<\bar\psi\psi>$ is seen by plotting the gauge ensemble average of the
second term in
(\ref{eq:pbp}) as a function of $m$ at a fixed $g=0.4\sqrt{\pi}$
for several lattice sizes. We focus on the small mass region in
Fig.\ref{fig:pbp1}. The data at $L=32$ show a rise in $<\bar\psi\psi>$ as
the mass is decreased. This divergence is due to an accumulation
of very small eigenvalues at larger $L$ as seen in the
histogram of the small non-zero eigenvalues in Fig.\ref{fig:hist1}.
Even though only $L=32$ shows a rise in $<\bar\psi\psi>$ at small masses,
an anomalous accumulation of very small eigenvalues is evident
at $L=24$ in Fig.\ref{fig:hist1}. The accumulation is not enough to
give a rise in $<\bar\psi\psi>$ on the $L=24$ lattice.

\begin{figure}
\epsfxsize = 0.9\textwidth
\caption{Plot of the distribution of the smallest eigenvalue scaled with
the volume and restricted to the zero topological sector
at $g=0.4\sqrt{\pi}$
for four different lattice sizes.}
\centerline{\epsffile{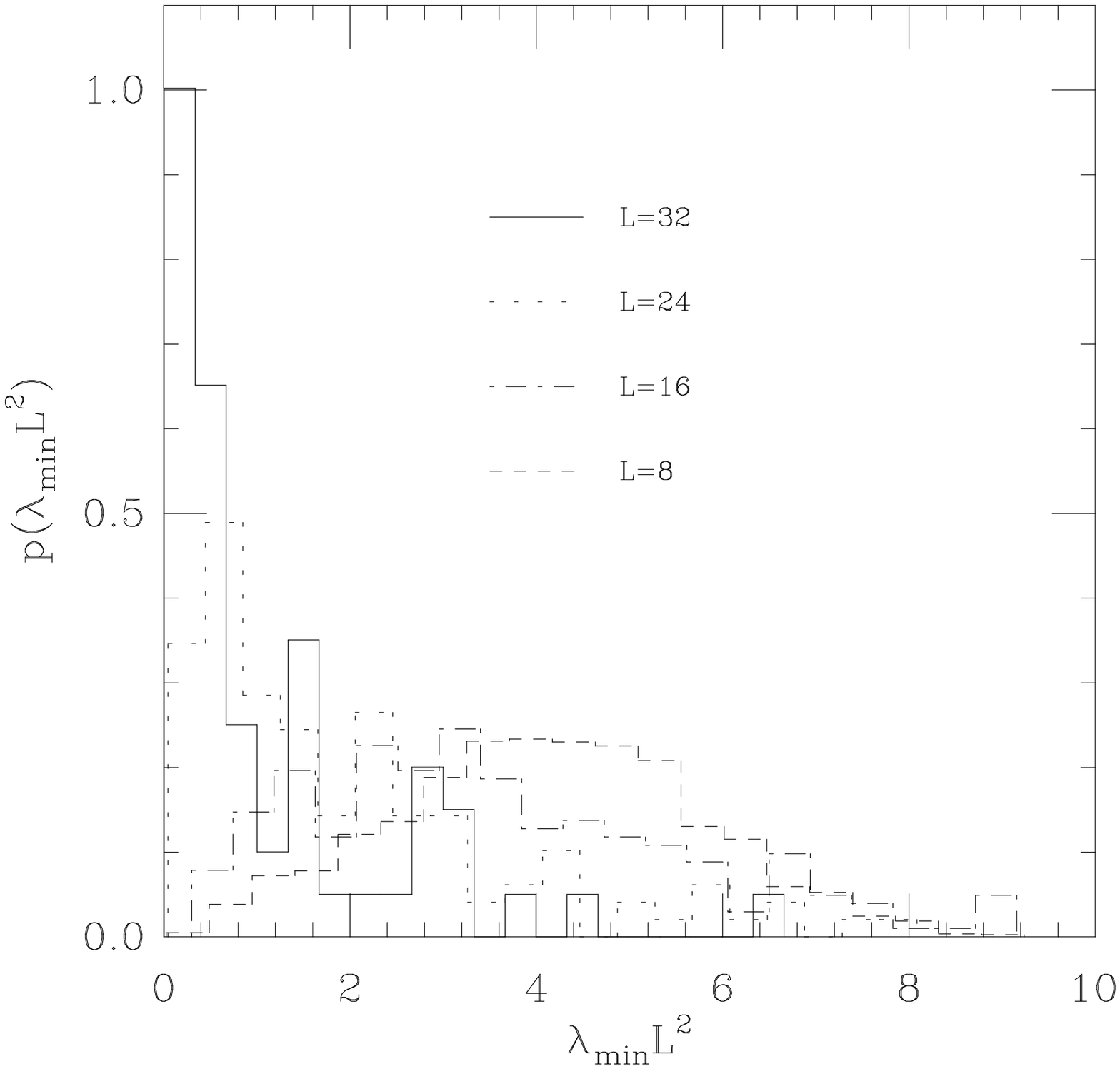}}
\label{fig:emin1}
\end{figure}

\begin{figure}
\epsfxsize = 0.9\textwidth
\caption{Ensemble average of the smallest eigenvalue in the
zero and unit topological sectors 
at $g=0.4\sqrt{\pi}$ as a function of the lattice size
and plotted to facilitate comparison with the simple functional forms
described in the text.
}
\centerline{\epsffile{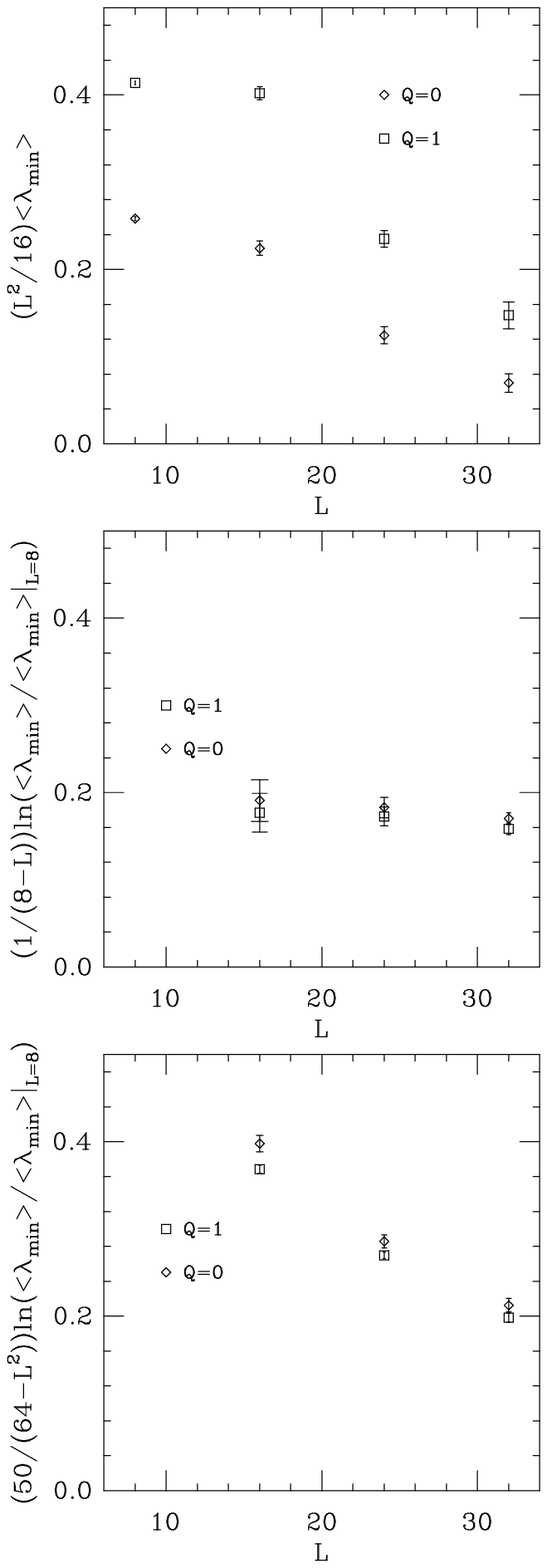}}
\label{fig:eminav}
\end{figure}

The smallest eigenvalue $\lambda_{\rm min}$ has to scale like $1/V$
for a finite value of the density of eigenvalues at zero $\rho(0)$ and
a finite value of $<\bar\psi\psi>$ in the massless limit.  In
Fig.\ref{fig:emin1}, we plot the histogram of $\lambda_{\rm min}L^2$
for the various ensembles in the zero topological sector.  We see that
there is no evidence for scaling in the distribution, whereas chiral
random matrix theory predicts a universal function of the form
${z\over 2}e^{-{z^2\over 4}}$ with $z=\Sigma L^2\lambda_{\rm min}$ and
$\Sigma$ the value of the chiral condensate.  
A previous analysis~\cite{Lang}
of the distribution of the low lying eigenvalues done at smaller physical
volume showed that the distribution did not fit the predictions of
unitary chiral random matrix theory~\cite{Wettig}. In Ref.~\cite{Lang}
this was attributed to
finite volume effects. In our case, the reason for the discrepancy is not small
volumes but a divergent chiral condensate.
%The average of the
%smallest eigenvalue is plotted as a function of $L$ in
%Fig.\ref{fig:eminav} in the $Q=0$ and $|Q|=1$ topological sectors.
%The data seem to fit a function of the form $e^{-\mu L}$ rather than $1/V$.  
Now consider the average of the smallest nonzero eigenvalue as a
function of $L$. Three simple functions motivated by heuristic physics 
arguments are $c/V$, $e^{-cL}$, and 
$e^{-cV}$. To determine which of these forms is closest to the data,
we have plotted in Fig.\ref{fig:eminav} 
$V \langle \lambda_{\rm min}\rangle/16$, 
$\ln(\langle \lambda_{\rm min} \rangle/
\langle \lambda_{\rm min} \rangle\mid_{L=8})/(8-L)$, and
$50 \ln(\langle \lambda_{\rm min} \rangle/\langle \lambda_{\rm min} \rangle
\mid_{L=8})/(64-V)$ verses $L$ for $Q=0$ and $|Q|=1$. 
(The normalizations are just for convenience.) To
the extent that one of these functions represents the data well,
the corresponding graph in Fig.\ref{fig:eminav} should be flat. 
Evidently $e^{-cL}$ is preferred.
Recall
that this is the form that appears in the argument for the lower bound
in \cite{Smilga} and in the spectrum from the artificial
configurations described in Section III.

\begin{figure}
\epsfxsize = 0.9\textwidth
\caption{Plot of $<\bar\psi\psi>/g$ with respect to $m/g$ at $gL/\sqrt{\pi}
=12.8$ on $L=24,28,32$. }
\centerline{\epsffile{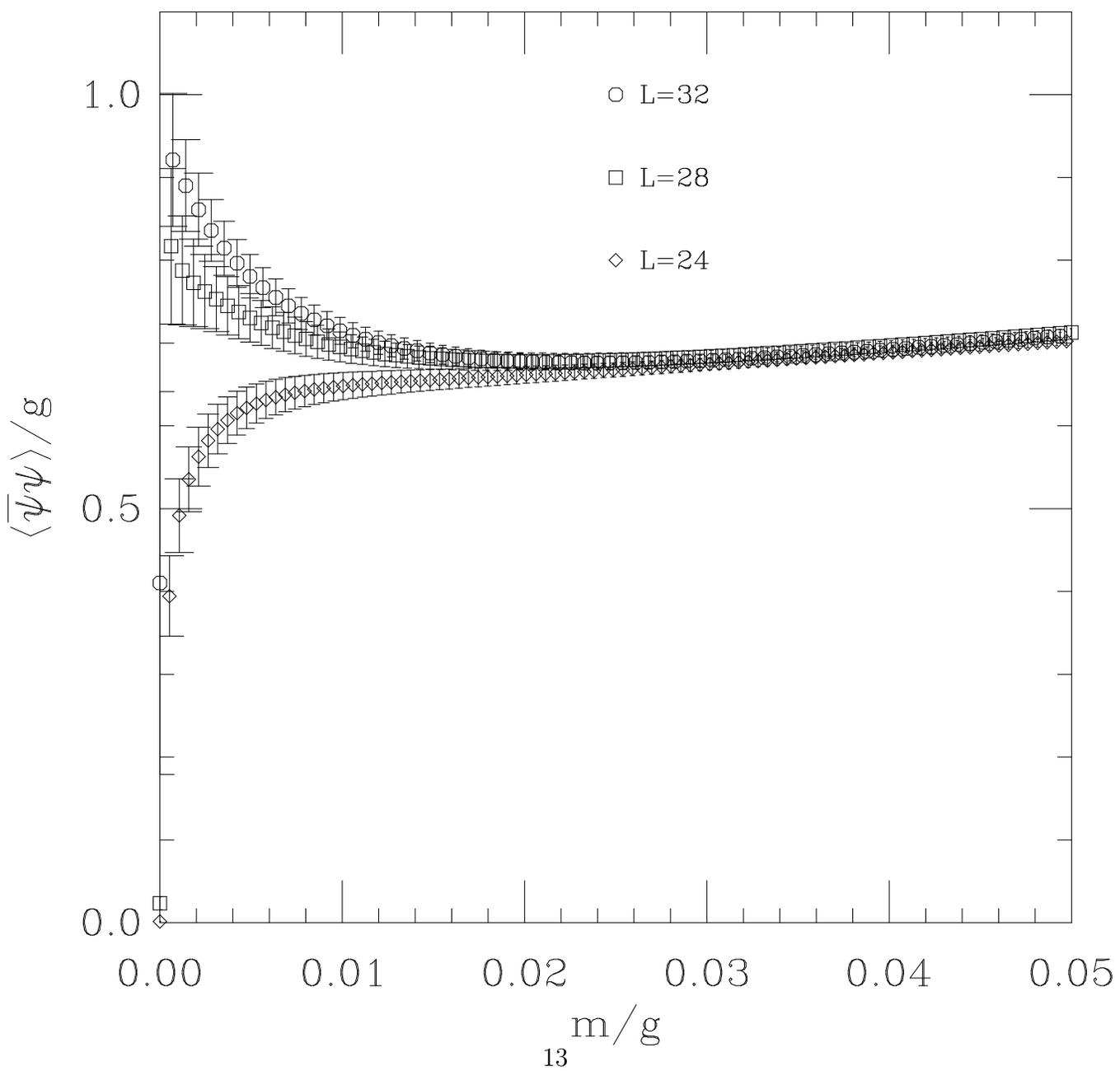}}
\label{fig:pbp2}
\end{figure}
\begin{figure}
\epsfxsize = 0.9\textwidth
\caption{Plot of $<\bar\psi\psi>/g$ with respect to $m/g$ 
at $L=32$ and four different couplings. }
\centerline{\epsffile{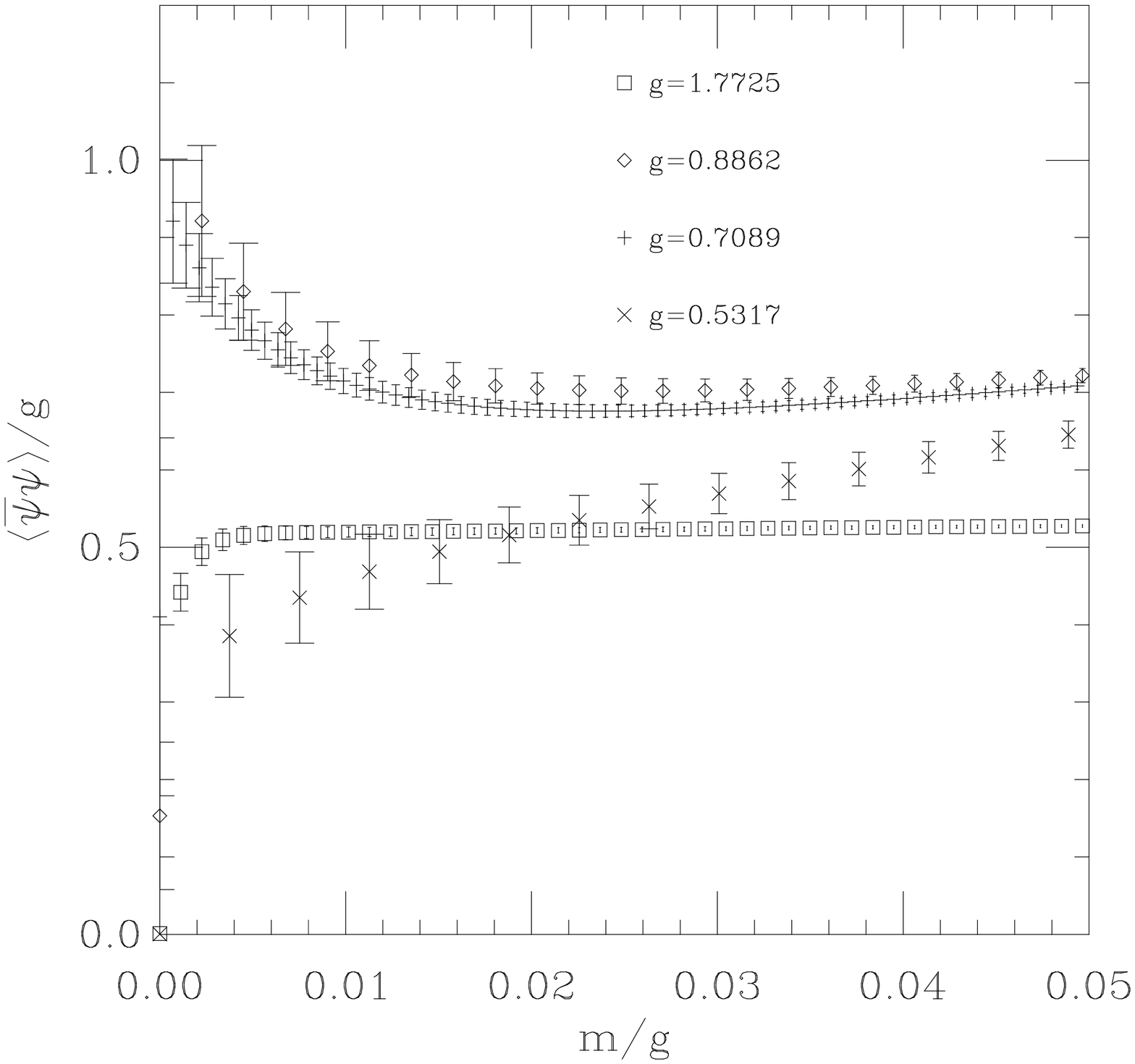}}
\label{fig:pbp3}
\end{figure}

Fig.\ref{fig:pbp1} shows that $L=32$ is needed at $g=0.4\sqrt{\pi}$ to
see the divergent behavior in the chiral condensate. This corresponds
to a physical volume of $gL/\sqrt{\pi}=12.8$. To study the effect of
lattice spacing, we compared this result with others obtained on
$L=24$ at $g/\sqrt{\pi}=0.4(32/24)$ and on $L=28$ at
$g/\sqrt{\pi}=0.4(32/28)$. These have the
same physical volume $gL/\sqrt{\pi}=12.8$ but are coarser lattices.
The comparison in Fig.\ref{fig:pbp2} shows that the divergence visible
on the $L=32$ lattice is not seen on the $L=24$ lattice. However, the $L=28$ 
data
follow the $L=32$ data to very small masses and into the region where the
condensate begins to grow.
We have used dimensionless quantities in this plot to facilitate
a proper comparison. The scaling behavior is good until $m/g$ gets small enough
to emphasize the very smallest eigenvalues, some of which are being distorted
on the coarser lattice. At smaller coupling and at the same
physical volume, the scaling behavior extends to smaller $m/g$.

To illustrate the point that at fixed $L$, $g$ can be neither too big nor too
small if the small $\lambda$ growth is to be seen, we have data from $L=32$ and
four couplings in Fig.\ref{fig:pbp3}. The smallest value of $g$, which corresponds to a
physical size of 9.6, shows no growth at all. The medium values at sizes 12.8
and 16 show the effect. (Note that the finite size effects between these two
are small.) The largest value of $g$ has size 32 but the gauge field there is
too rough, and the small $\lambda$ peak is gone.

\section{Conclusions}

We have shown that for small coupling and large volume, a small
eigenvalue peak appears in the spectral density of the overlap Dirac
operator and in $<\bar\psi\psi>$. This is strong evidence for the
predictions~\cite{Smilga,Durr} that these quantities diverge in the
infinite volume limit of the quenched Schwinger model. There is some
evidence that the divergence could be as strong as $e^{cgL}$, but the
lattice sizes are insufficient to provide evidence for the stronger
$e^{dg^2L^2}$ predictions. Similarly there is limited evidence that
the would-be-zero modes of subregions of the lattice can provide a
physical understanding of the results. However, a definite test of
that model also awaits data from larger lattices.

The results in this paper
clearly point the direction for further work. The calculations
should be extended to larger lattices so that there are several sizes
showing the small $\lambda$ growth of the spectral density and 
$<\bar\psi\psi>$. With that, it would be possible to study the volume 
dependence
of the small $\lambda$ peaks and test in more detail the theoretical
expectations. 

Although the quenched Schwinger model is quite some distance from full
four-dimensional QCD, results from it help to map out the range of territory
available to massless fermions responding to a gauge field.

\acknowledgements

R.N. would like to thank Urs Heller for general discussions on the
quenched approximation and Herbert Neuberger for some discussions on
the quenched Schwinger model.

\end{document}